\documentclass[aps,prb,reprint,amsmath,amssymb,floatfix,superscriptaddress,nofootinbib]{revtex4-2}
\usepackage{graphicx}
\usepackage{dcolumn}
\usepackage{bm}
\usepackage{xcolor}
\usepackage{braket}
\usepackage[normalem]{ulem}
\usepackage{cancel}
\usepackage{mwe}

\begin{document}
\title{Symmetry-Projected Spin-AGP Methods Applied to Spin Systems}

\author{Zhiyuan Liu}

\affiliation{Department of Physics and Astronomy, Rice University, Houston, TX 77005-1892, USA}

\author{Thomas M. Henderson}
\affiliation{Department of Physics and Astronomy, Rice University, Houston, TX 77005-1892, USA}
\affiliation{Department of Chemistry, Rice University, Houston, TX 77005-1892, USA}

\author{Gustavo E. Scuseria}
\affiliation{Department of Physics and Astronomy, Rice University, Houston, TX 77005-1892, USA}
\affiliation{Department of Chemistry, Rice University, Houston, TX 77005-1892, USA}

\date{\today}

\begin{abstract}
Symmetry-projected wave function methods capture static correlation by breaking and restoring the symmetries of a system. In this article, we present the symmetry-projected spin antisymmetrized geminal power (spin-AGP) state projected onto space group symmetry as well as complex conjugation, spin-flip, and time-reversal symmetries. The method is benchmarked on the 1D XXZ model and 2D $\mathrm{J_1-J_2}$ model with square and triangular lattices. Our results indicate that symmetry projection methods provide a powerful tool for frustrated spin systems.

\end{abstract}

\maketitle

\section{Introduction}

 Within the condensed matter physics community, spin model Hamiltonians have long served as a primary framework to study frustrated quantum magnetism \cite{mikeska2004one,science.abm2295}. With a few exceptions \cite{KITAEV20062}, spin lattice models beyond one dimension are not exactly solvable, requiring the use of approximate numerical approaches instead. Analogously to the Hartree--Fock method used in the electronic structure theory of fermionic systems, spin wave functions based on a single spin configuration are insufficient for strongly correlated systems. Recent research has shown that the spin antisymmetrized geminal power (spin-AGP) method, which extends the AGP \cite{Coleman1965,ring2004nuclear} concept from systems of paired fermions to spin systems, may provide a good reference state for spin systems \cite{PhysRevB.108.085136}. However, it is notable that, despite yielding some good results for one-dimensional (1D) systems, spin-AGP and correlated methods based on it are less accurate for lattices with more complicated geometry. 

In spin lattice models, the spin-AGP is usually forced to obey the same spatial symmetry as the lattice \cite{PhysRevB.108.085136}, severely compromising the flexibility of the ansatz. In this paper, we propose to solve this problem by deliberately breaking spatial symmetry and projectively restoring it, as similar symmetry projection methods for fermionic systems allow a description of many important strong correlations at low computational cost \cite{https://doi.org/10.1002/qua.560080515,10.1063/1.4705280,10.1063/1.4991020,10.1063/5.0080165}.  

In addition, we will see later in this article that the projection of the space group symmetry in spin-AGP is often accompanied by complex conjugation and time reversal symmetry breaking, so we projectively restore these symmetries as well. Notably, complex conjugation and time-reversal symmetries, together with spin-flip symmetry, form a closed set. The symmetry projection of these three has yielded favorable results for fermionic systems \cite{doi:10.1021/acs.jpca.4c03127}, particularly in conjunction with the projection of point group symmetry of a molecule.

 In this article, we review spin-AGP methods in Sec.~\ref{Sec.sAGP}. In Sec.~\ref{Sec.space} and Sec.~\ref{Sec.complex}, we present space group symmetry and complex conjugation projection, spin-flip, and time-reversal symmetry projection for the spin-lattice. These methods will be applied to the 1D XXZ ring and the the two-dimensional (2D) $\mathrm{J_1-J_2}$ model with square and triangular lattices in Sec.~\ref{Sec.Result}.

\section{Theory}
\subsection{Spin-AGP}
\label{Sec.sAGP}
 The antisymmetrized geminal power (AGP) is a mean-field method of seniority zero electron pairs. The spin-AGP state is an extension of the AGP applied to spin-1/2 systems. It is a wave function defined by the geminal operator
\begin{equation}
    \Gamma^\dagger=\sum_{p=1}^M \eta_p \, S_p^+,
    \label{Gamma}
\end{equation}
where $M$ is the number of sites, $S_p^+$ is the spin-1/2 raising operator on the spin at site $p$ and $\eta_p$ is the AGP geminal coefficient.  The spin AGP wave function is then 
\begin{equation}
    |\mathrm{sAGP}\rangle = \frac{1}{N!} \, \left(\Gamma^\dagger\right)^N \ket{\Downarrow},
    \label{sAGP}
\end{equation}
where $N$ is the number of $\uparrow$ spins and $\ket{\Downarrow}$ is the product state of $\downarrow$ spins on all sites, satisfying
\begin{equation}
    S_p^-\ket{\Downarrow}=0,~\forall p.
\end{equation}

It is evident that the spin-AGP wave function is an eigenstate of the global $S^z$ with
\begin{equation}
  S^z|\mathrm{sAGP}\rangle = (N - \frac{M}{2})|\mathrm{sAGP}\rangle.
\end{equation}
At half-filling ($N = M/2$), we have $\braket{S^z}=0$, and the spin-AGP wave function is magnetically neutral.

It should be noted that spin-AGP is equivalent to the $S^z$-projected spin BCS (sBCS) state, 
\begin{equation}
|\mathrm{sAGP}\rangle =P_{S_z}\left\vert \mathrm{sBCS}\right\rangle
\end{equation}
where sBCS is defined as
\begin{align}
\left\vert \mathrm{sBCS}\right\rangle 
&=\mathrm{exp}(\Gamma^\dagger)\ket{\Downarrow}\nonumber\\
&=\prod\limits_{p=1}^{M}\left( 1+\eta _{p}S_{p}^{+}\right)
\ket{\Downarrow}
\end{align}

The spin-AGP state can be optimized variationally with $\mathcal{O}(M^3)$ cost \cite{PhysRevB.108.085136,khamoshi2019efficient, 10.1063/5.0156124}. Correlation methods are also feasible for spin-AGP \cite{PhysRevB.108.085136,henderson2019geminal,henderson2020correlating,dutta2020geminal,dutta2021construction,khamoshi2021exploring}. 

Throughout the remainder of this article, we will simplify our terminology by referring to spin-AGP simply as AGP, as all discussions pertain to spin systems, making this distinction unnecessary.

\subsection{Space group symmetry projection}
\label{Sec.space}
Spin-lattice models usually exhibit space group symmetry. A mean-field optimized AGP, however, may spontaneously break space group symmetry even though the exact ground state observes the symmetry. For example, the 1D XXZ Hamiltonian (see below in Sec.~\ref{sec.XXZ}) with periodic boundary conditions and $M$ sites has $D_M$ space group symmetry.  Its ground state for finite $M$, therefore, transforms according to an irreducible representation of $D_M$, meaning it is symmetry adapted. However, the mean-field optimized AGP breaks space group symmetry for $\Delta\gtrsim1$ and $\Delta\lesssim-1$. To correctly predict the properties of the system, it is vital to restore space group symmetry of the mean-field optimized state by a symmetry projection:
\begin{equation}
      |\mathrm{SG-AGP}\rangle = P_{SG} |\mathrm{AGP}\rangle,
    \label{SGAGP}
\end{equation}
where $|\mathrm{SG-AGP}\rangle$ is the symmetry-projected AGP state and $P_{SG}$ is the space group symmetry projection operator.

The space group projection operator can be written as 
 \begin{equation}
    P_{SG} = \sum_G \mathrm{e}^{\mathrm{i} \theta_G} G,
    \label{SG_p2}
\end{equation}
where $G$ runs over all operators in the space group.  For symmorphic space groups, which are applicable to all lattices examined in this paper, space group symmetry can be decomposed into lattice momentum symmetry (i.e. translational symmetry) and point group symmetry, and the space group projection operator factorizes:
\begin{subequations}
    \begin{align}
    P_{SG}(\mathbf{q},k) &= P_{LM}(\mathbf{q}) \times P_{PG}(k),\\
      P_{LM}(\mathbf{q}) &= \sum_{\mathbf{r}} e^{ \mathrm{i} \mathbf{q}\cdot\mathbf{r}}~ T_{\mathbf{r}},\\
      P_{PG}(k) &= \sum_l \chi_k(l) R_l,\
    \end{align}
    \label{SG-projection}
\end{subequations}
$P_{LM}$ and $P_{PG}$ are the lattice momentum symmetry projection operator and point group symmetry projection operator, respectively. $T_{\mathbf{r}}$ is the translational operator with $\mathbf{q}$ being the lattice momentum projected onto. $R_l$ is a symmetry operator of the lattice point group and $\chi_k(l)$ is the character of the operator $R_l$ in irreducible representation $k$. 

Note that space group operators in practice permute sites of the lattice.  Therefore, space group operators acting on AGP produce another AGP whose geminal coefficients $\eta$ are permuted:
\begin{equation}
    G|\mathrm{AGP}(\eta)\rangle = |\mathrm{AGP}(\eta_{G})\rangle.
    \label{AGP_perm}
\end{equation}
 For example, if the symmetry operator $G$ is the translation operator that translates site $i$ into $i+1$, and $\{\eta_1,\eta_2,\cdots,\eta_M\}$ is the geminal coefficient of the original AGP, we have $|\mathrm{AGP}(\eta_{G})\rangle=G|\mathrm{AGP}(\eta)\rangle=G |\eta_1,\eta_2,\cdots,\eta_M\rangle =|\eta_2,\eta_3,\cdots,\eta_{1}\rangle$.
 
Combining Eqn.~\ref{SG_p2} and \ref{AGP_perm}, the space group symmetry projection on AGP is thus achieved by a linear combination (LC) of AGPs:
\begin{equation}
      |\mathrm{SG-AGP}\rangle =\sum_{G} e^{\mathrm{i}\theta_G} |\mathrm{AGP(\eta_{G})}\rangle.
    \label{SGAGP_LC}
\end{equation}

Noting that $P_{SG}$ commutes with the Hamiltonian $H$ and $P_{SG}^2=P_{SG}$, we have

    \begin{align}
        E_{SG-AGP}&=\frac{\langle \mathrm{AGP}|P_{SG}H P_{SG}|\mathrm{AGP}\rangle}{\langle \mathrm{AGP}|P_{SG}P_{SG}|\mathrm{AGP}\rangle}\nonumber\\
        &=\frac{\langle \mathrm{AGP}|H P_{SG}|\mathrm{AGP}\rangle}{\langle \mathrm{AGP}|P_{SG}|\mathrm{AGP}\rangle}\nonumber\\
        &=\frac{\sum_{G} e^{i\theta_G} \langle \mathrm{AGP}|H|\mathrm{AGP(\eta_{G})}\rangle}{\sum_{G} e^{i\theta_G} \langle \mathrm{AGP}|\mathrm{AGP(\eta_{G})}\rangle}.
        \label{E_SGAGP}
    \end{align}

We note that the bra state can be simplified to the original AGP without permutation. Only the ket state need be explicitly projected, and therefore represented as a linear combination of non-orthogonal AGPs (LC-AGP). Since the number of space group symmetry operators, and hence the number of AGPs required for this LC-AGP, is typically $\mathcal{O}(M)$, where $M$ is the number of spin sites, the cost of space group symmetry-projected AGP is one factor of $M$ larger than that of the mean-field optimized AGP, making it $\mathcal{O}(M^4)$. This is in contrast to a more general LC-AGP--style wave function, for which one needs to evaluate Hamiltonian overlaps between $\mathcal{O}(M^2)$ pairs of AGPs, at an $\mathcal{O}(M^5)$ computational cost.

\subsection{Complex Conjugate, Spin-Flip, and Time-Reversal Symmetry}
\label{Sec.complex}
As we shall see below, variationally optimizing the space group projected AGP generally requires complex $\eta$ parameters, meaning that the SG-AGP typically breaks complex conjugation and time reversal symmetries spontaneously.  We can take advantage of this fact to projectively restore these symmetries as well \cite{doi:10.1021/acs.jpca.4c03127}, which provides additional variational flexibility.

The complex conjugation projector can be written as 
\begin{equation}
      P_K = 1 + \mathrm{e}^{i\theta}K.
    \label{P_K_ori}
\end{equation}
where the complex conjugation operator $K$ is an anti-unitary operator and satisfies
\begin{subequations}
    \begin{align}
      &\langle K \psi |\phi\rangle =\langle \psi |K^\dagger \phi\rangle^*,\\
        & K K^\dagger = K^\dagger K =KK =1.
    \end{align}
    \label{K}
\end{subequations}
In practice, the phase $\theta$ present in Eqn.~\ref{P_K_ori} can be incorporated into the projected state as a global phase \cite{doi:10.1021/acs.jpca.4c03127}, thereby simplifying the projector to
\begin{equation}
      P_K = 1 + K.
    \label{P_K}
\end{equation}

Spin-flip projection takes the form
\begin{equation}
    P_F = 1\pm F.
    \label{P_F}
\end{equation}
This projection is also known as half-spin projection \cite{smeyers1971etude,SMEYERS2000253}, and the $+$ and $-$ signs in Eqn.~\ref{P_F} correspond to projection onto even and odd spin parity, respectively.  In other words, the $+$ sign projects onto states with even $s$ and the $-$ sign projects onto states with odd $s$, where $s$ is the spin quantum number, with eigenstates of the $S^2$ operator having eigenvalue $s (s+1)$.

For an AGP state with geminal coefficients $\eta$, it can be shown that $K$ or $F$ on the AGP state results in another AGP whose geminal coefficients are the complex conjugate or negative reciprocal of the original AGP, respectively:
\begin{subequations}
    \begin{align}
        &K|\mathrm{AGP(\eta)}\rangle = |\mathrm{AGP(\eta^*)}\rangle ,\\
        &F|\mathrm{AGP(\eta)}\rangle = |\mathrm{AGP(-1/\eta)}\rangle.
    \end{align}
    \label{KFAGP}
\end{subequations}
We apply the complex conjugate and spin-flip projections on the AGP states in addition to the space group symmetry projection.

The complex conjugation operator, together with the spin-flip operator $F=e^{\mathrm{i}\pi S_y}$ and the time-reversal operator $\mathcal{T}=FK$, forms a closed set:
\begin{subequations}
    \begin{align}
        &\mathcal{T}=FK=KF,\\
        &F=K\mathcal{T}=\mathcal{T}K,\\
        &K=-F\mathcal{T}=-\mathcal{T}F.
    \end{align}
    \label{FTK}
\end{subequations}
Because this set is closed, time-reversal symmetry is automatically projected once we apply the complex conjugation and spin-flip projectors.

\section{Applications}
\label{Sec.Result}
In this article, space group projected AGP methods are tested on various one-dimensional and two-dimensional lattices. The space group projected AGP ({Eqn.~\ref{E_SGAGP}}) is optimized with the variation after projection approach, which means that we directly optimize the energy of the projected state with respect to the parameters (\textit{i.e.} the $\eta$ values) of the unprojected one.

\subsection{One-dimensional XXZ model}
\label{sec.XXZ}
We take the antiferromagnetic XXZ model as our benchmark model for one-dimensional systems. The XXZ Hamiltonian is

    \begin{align}
    \label{XXZ} 
      H_{\mathrm{XXZ}}
      &= \, \sum_{\langle pq \rangle} \left(S_p^x \, S_q^x + S_p^y \, S_q^y + \Delta \, S_p^z \, S_q^z\right)
    \nonumber\\
      &= \, \sum_{\langle pq \rangle} \left[\frac{1}{2} \, \left(S_p^+ \, S_q^- + S_p^- \, S_q^+\right) + \Delta \, S_p^z \, S_q^z\right], 
    \end{align}
where $p$ and $q$ index lattice sites and the notation $\langle pq \rangle$ restricts
the summation over $p$ and $q$ to nearest neighbors. The system exhibits a N\'eel antiferromagnetic phase for $\Delta \gtrsim 1$, and a ferromagnetic phase for $\Delta \lesssim -1$.  In the interval region $|\Delta| \lesssim 1$, the system is in the XY phase characterized by gapless excitations and long-range correlations \cite{mikeska2004one,bishop1992coupled,bishop1996coupled}.

\begin{figure}[b]
\includegraphics[width=\columnwidth]{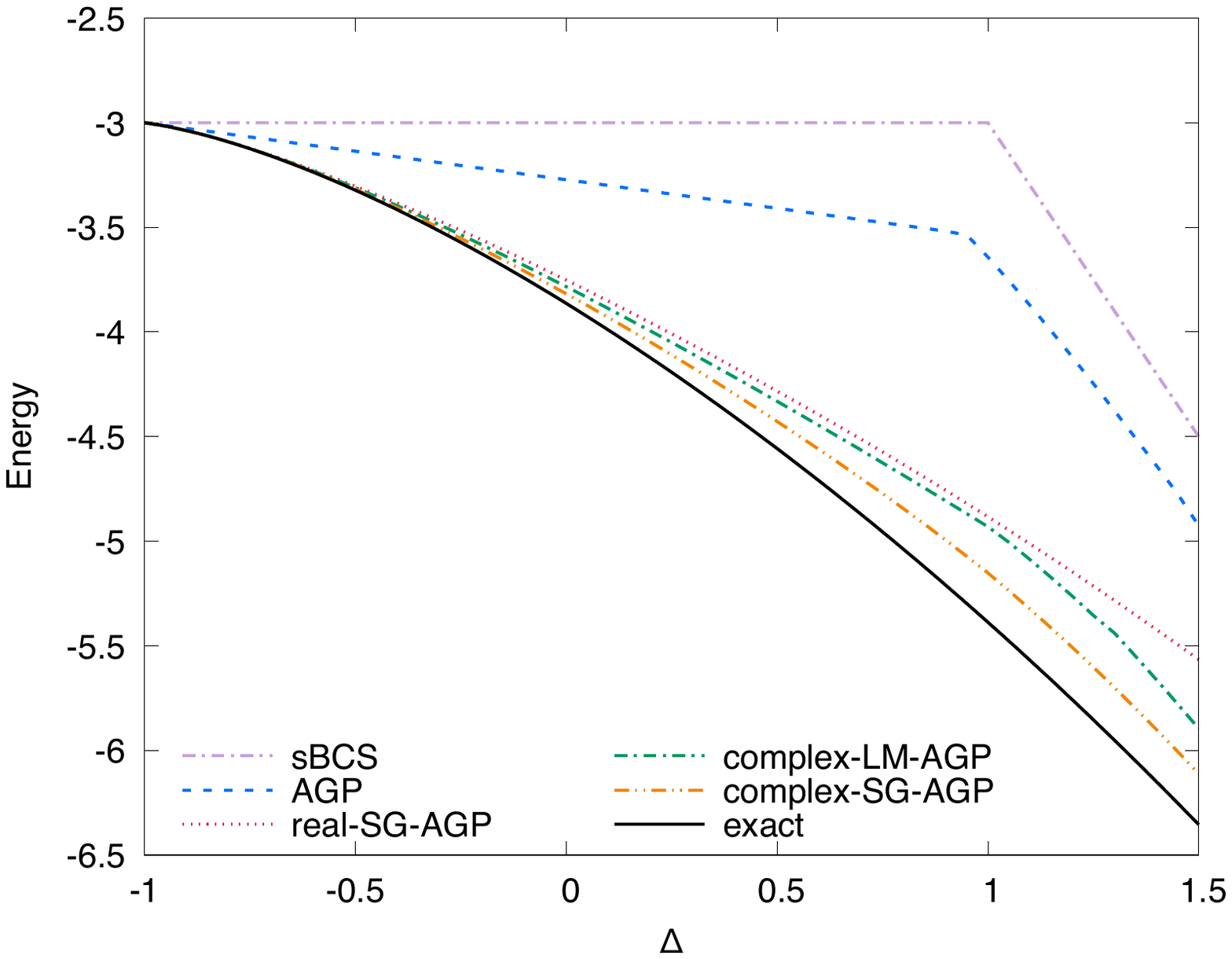}
\caption{Energy of the 12-site 1D XXZ model with periodic boundary conditions by sBCS, AGP and space group projected AGP methods. SG-AGP with real $\eta$ is denoted as real-SG-AGP, while complex-LM-AGP and complex-SG-AGP represent LM-AGP and SG-AGP with complex $\eta$, respectively. These methods are compared with mean-field optimized AGP.}
\label{fig:E_1DXXZ_1}
\end{figure}

For the 1D XXZ model with infinite length, the lattice has $D_\infty$ space group symmetry. For an $M$-site finite lattice with periodic boundary conditions, the symmetry is reduced to $D_M$. This symmetry can be decomposed into translation/rotation operations along the ring and mirroring of the lattice. The space group contains 2$M$ symmetry operations and the projected AGP can be written as

    \begin{align}
        |SG-AGP\rangle &=  P_{SG}|\eta_1,\eta_2,\cdots,\eta_M\rangle\nonumber\\
        &= \sum_{i=0}^{M-1}|\eta_{i+1},\eta_{i+2},\cdots,\eta_i\rangle\nonumber\\
        &+ \sum_{i=0}^{M-1}|\eta_{i-1},\eta_{i-2},\cdots,\eta_i\rangle,
        \label{1DSG}
    \end{align}
which is a linear combination of 2$M$ AGP states.  Note that the first term on the right-hand-side corresponds to lattice momentum projection onto $\mathbf{k} = \boldsymbol{0}$; the addition of the second summation adds the point-group projection.

\begin{figure}[t]
    \includegraphics[width=\columnwidth]{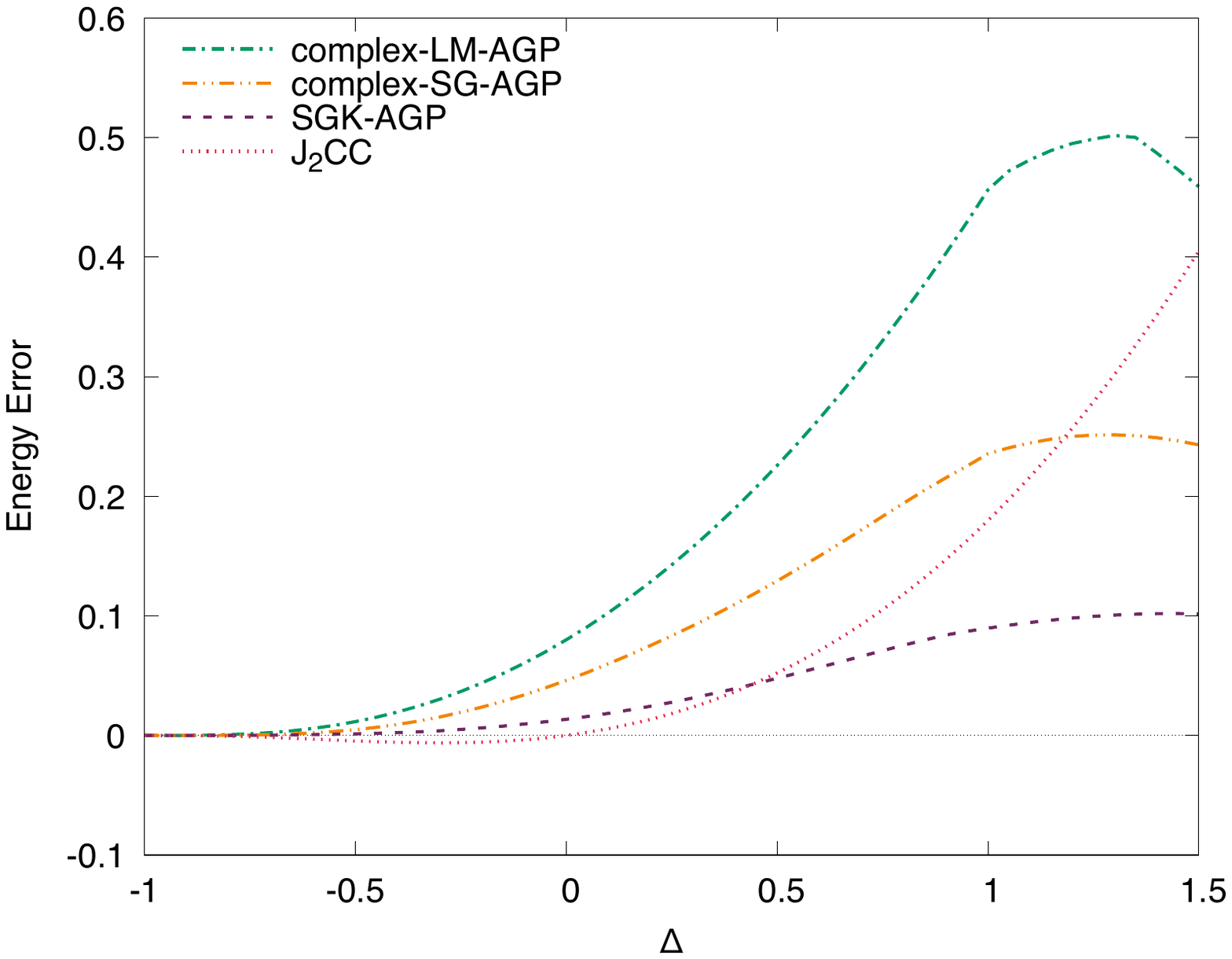}
    \caption{Energy error of real-SG-AGP, complex-LM-AGP, complex-SG-AGP and SGK-AGP on the 12-site 1D XXZ model with periodic boundary conditions. These methods are compared with J$_2$CC.  Note that SGK-AGP denotes the combination of space group and complex conjugation projection.}
    \label{fig:E_1DXXZ_3}
\end{figure}

The space group projected AGP is optimized with both real and complex geminal coefficient $\eta$. The energies of the 12-site XXZ chain with periodic boundary conditions ($M$=12) are shown in Fig.~\ref{fig:E_1DXXZ_1}. The result of lattice momentum projected AGP (LM-AGP), which projects lattice momentum symmetry but not point group symmetry, with complex $\eta$ is also included. The results are compared with the mean-field optimized sBCS and AGP methods. Projecting $S^z$ to convert sBCS into AGP provides a noticeable but not large improvement to the total energy. However, the improvement upon incorporating space group projection as well is much larger. Furthermore, with complex $\eta$, the energy result is much superior to that of real $\eta$ at the cost of the complex conjugation symmetry. The errors of complex SG-AGP methods are shown in Fig.~\ref{fig:E_1DXXZ_3} and are compared with J$_2$ coupled cluster (J$_2$CC) \cite{PhysRev.108.1175}, a correlated method based on a single AGP state. We note that the symmetry projection makes a huge difference in the energy results, and the complex-SG-AGP, as a symmetry-projected mean-field method, already produces a result comparable to correlated methods.

To overcome the issue of breaking complex conjugation symmetry by making $\eta$ complex, the projection of complex conjugation symmetry is implemented together with the space group symmetry projection, denoted as SGK-AGP in Fig.~\ref{fig:E_1DXXZ_3}.  This not only restores the correct complex conjugation symmetry but also significantly improves the energy.  Because SG-AGP appears to yield significantly better results with complex $\eta$ than with real $\eta$, all space group projected AGP results discussed in this manuscript are optimized using complex $\eta$ unless otherwise stated.

Of course the energy is not the only quantity of interest. Correlation functions, in particular, are important and can help us understand the nature of magnetism. A single AGP is generally inadequate here, as it provides exact correlation functions at $\Delta = -1$ but predicts that the correlation function is independent of $\Delta$ for $|\Delta| \lesssim 1$.

\begin{figure}[t]
    \includegraphics[width=\columnwidth]{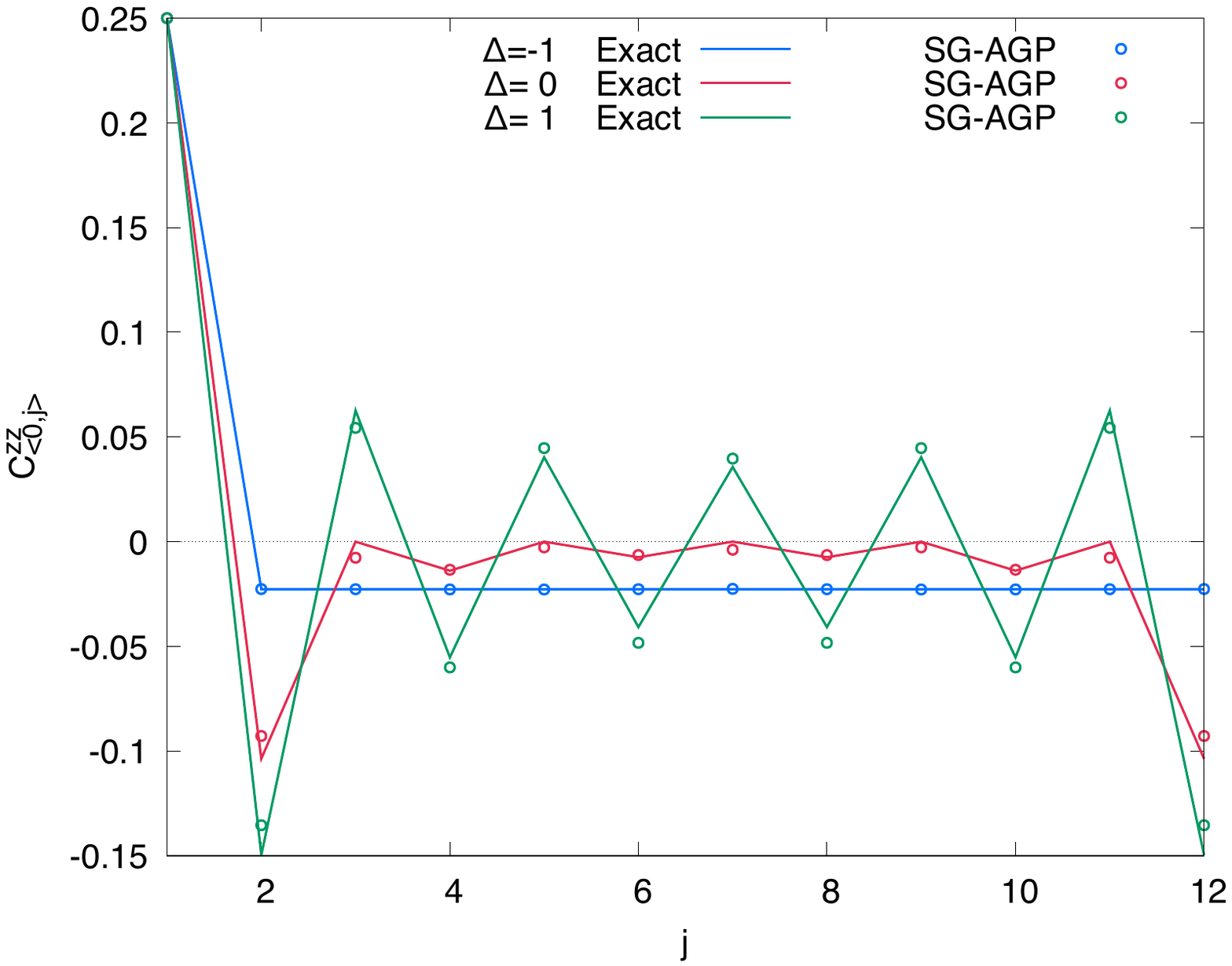}
    \caption{$S^z$-$S^z$ correlation functions for the 12-site XXZ model with PBC, compared to those from SG-AGP with exact results. The lines are for exact results, and the circles are SG-AGP. Different colors represent different $\Delta$ values.}
    \label{fig:E_1DXXZcorr}
\end{figure}

The $S^z-S^z$ spin-spin correlation is defined as 
\begin{subequations}
    \label{Sz_corr}
    \begin{align}
        C^{zz}_{<i,j>} = {\langle S_i^zS_j^z \rangle},
    \end{align}
\end{subequations}
where $i$ and $j$ are indices for sites. The results are shown in Fig.~\ref{fig:E_1DXXZcorr}. It can be seen that the spin-spin correlation given by SG-AGP is quite close to exact, and as $\Delta$ goes from $-1$ to $1$, the system approaches a N\'eel structure.

\begin{figure}[t]
\includegraphics[width=\columnwidth]{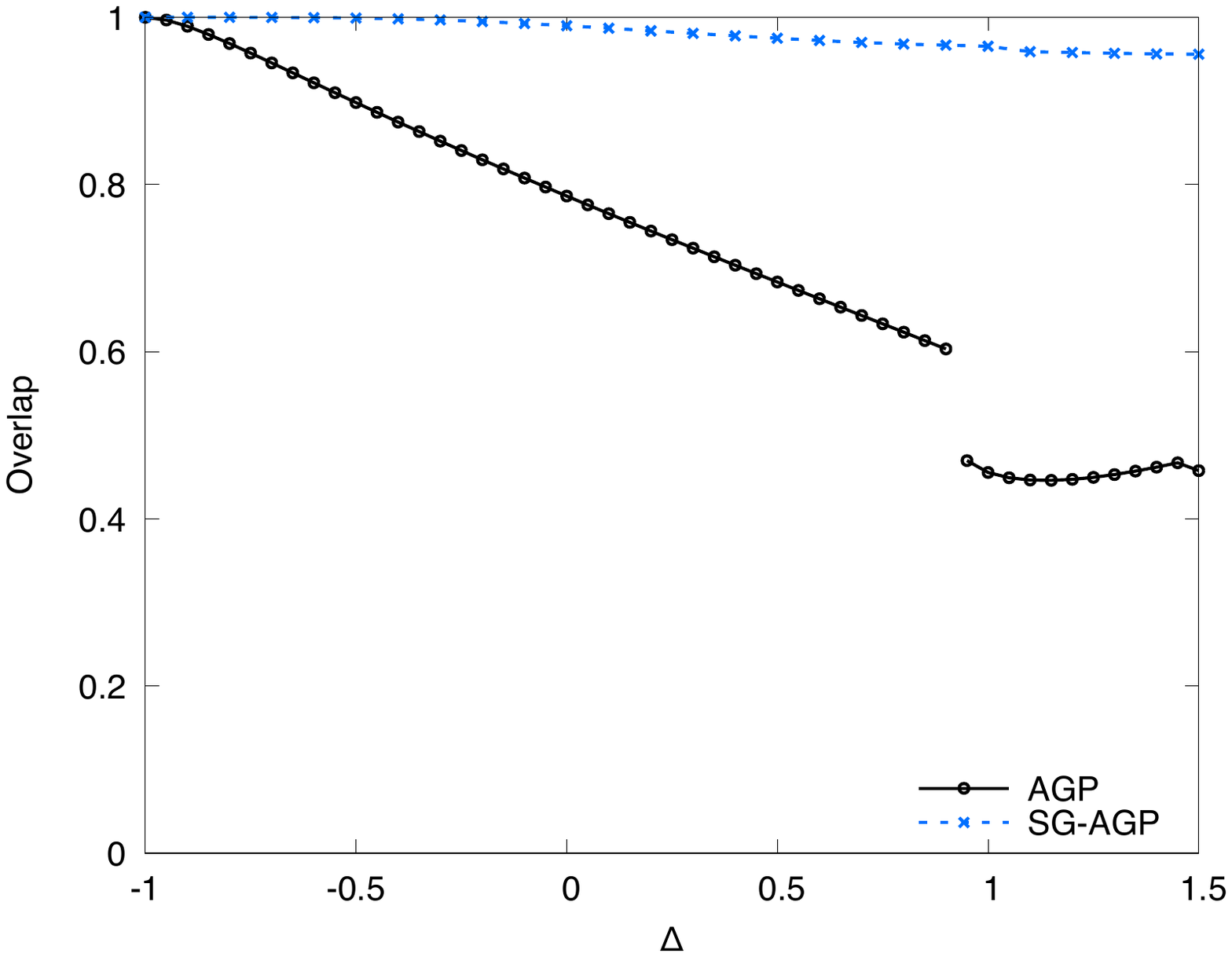}
\caption{AGP and SG-AGP overlap with full CI (FCI). The discontinuity in the AGP curve is because AGP has two solutions: one for $\Delta \lesssim 0.95$ and another for $\Delta \gtrsim 0.95$.}
\label{fig:1DXXZ_overlap}
\end{figure}

Figure~\ref{fig:1DXXZ_overlap} displays the overlap between the SG-AGP and the exact solution (FCI). This overlap is compared with that of AGP. While the AGP has relatively small overlap with the exact result except near $\Delta = -1$, the overlap of the SG-AGP and the exact solution is greater than 0.95 for all $\Delta$.  This indicates that SG-AGP captures most of the exact wave function, and suggests that the remaining correlations atop SG-AGP, unlike those atop AGP, are small.

\subsection{Two-dimensional $\mathrm{J_1-J_2}$ model}

In one-dimensional systems, the space group is merely reflection and translation confined to a single direction. However, in two-dimensional systems, the lattice structures become more sophisticated, resulting in significantly more diverse space groups. Therefore, space group projection is presumably more important for 2D systems. In this section, we will test our space group projected AGP methods on the $\mathrm{J_1-J_2}$ model with different 2D lattices.

The Hamiltonian of the $\mathrm{J_1-J_2}$ model is
\begin{equation}
  H_\mathrm{J_1-J_2}
  = \mathrm{J_1} \, \sum_{\langle pq \rangle} \left(\vec{S_p}\cdot\vec{S_q}\right)
  + \mathrm{J_2} \, \sum_{\langle\langle pq \rangle\rangle} \left(\vec{S_p}\cdot\vec{S_q}\right),
\end{equation}
where  $\langle pq \rangle$ and $\langle\langle pq \rangle\rangle$ denote sites $p$ and $q$ being nearest neighbors and next-nearest neighbors respectively. We take $\mathrm{J_1}=1$, and the competition between nearest neighbors and next-nearest neighbors is only dependent on $\mathrm{J_2}$. The space group projected AGP methods are tested on the square and triangular lattices, both of which show rich physics and have spin-liquid \cite{RevModPhys.89.025003} phases for intermediate $\mathrm{J_2}$. 

\begin{figure}
    \includegraphics[width=0.4\columnwidth]{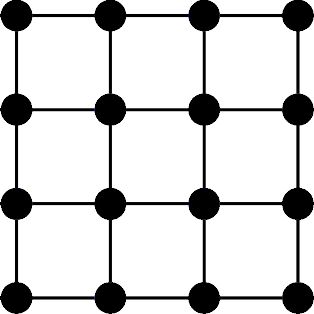}
    \hfill
    \includegraphics[width=0.5\columnwidth]{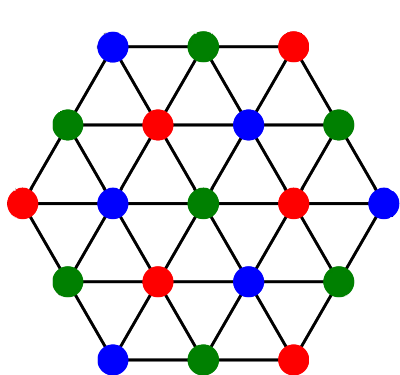}
\\
    \includegraphics[width=0.38\columnwidth]{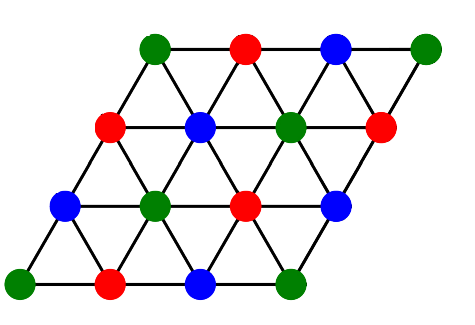}
    \hfill
    \includegraphics[width=0.57\columnwidth]{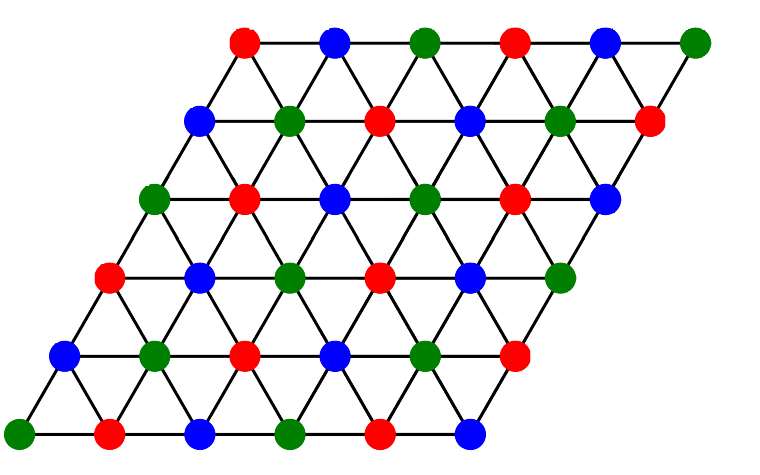}

    \caption{\label{Fig:twoDLat}Top left: Two-dimensional square lattice. Top right: the 12-site triangular lattice with a hexagon cell. The cell has periodic boundary conditions, and the top side, top right side, and top left side are the same as the the bottom side, bottom left side, and bottom right side, respectively. Bottom left:16-site triangular lattice with a rhombus cell. The lattice cannot accommodate the 120\textdegree~phase of the $\mathrm{J_1-J_2}$ model (see Sec.~\ref{Tri_result}). Bottom right: 36-site triangular lattice with a rhombus cell.  The color-coding on the triangular lattices helps clarify which sites are equivalent in the 120\textdegree~phase.}
  \end{figure}

\subsubsection{Square lattice}

A 2D square lattice is shown in the top left of Fig.~\ref{Fig:twoDLat}. The solid line connects the nearest neighbors, while the second-nearest neighbors differ by one site each in the horizontal and vertical directions. The system has the $p4m$ space group. The associated $D_4$ point group symmetry has eight symmetry operations, including rotation by $\pi/2, \pi, 3\pi/2$, and mirroring. The lattice momentum symmetry is achieved by the translation along the two lattice vectors. Periodic boundary conditions are enforced in both directions, making the lattice a torus. The SG-AGP of an $L\times L$ 2D square lattice is then a linear combination of $8L^2$ AGP states.

For the $\mathrm{J_1-J_2}$ model in the thermodynamic limit, when $\mathrm{J_2} \lesssim 0.45$, the system exhibits a checkerboard N{\'e}el order where all spins are antiparallel to their nearest neighbors. For $\mathrm{J_2} \gtrsim 0.61$, the system is in a well-established striped order with spins parallel in the same column (or row) but antiparallel between columns (or rows) \cite{LIU20221034}. For $\mathrm{J_2} \approx 0.5$, however, the system is in a highly frustrated phase. The ground state is under debate and possible candidates include the plaquette valence-bond state \cite{zhitomirsky1996valence}, the stripe valence-bond state \cite{sachdev1990bond}, or the gapless spin liquid state \cite{capriotti2001resonating}.

\begin{figure}
    \includegraphics[width=\columnwidth]{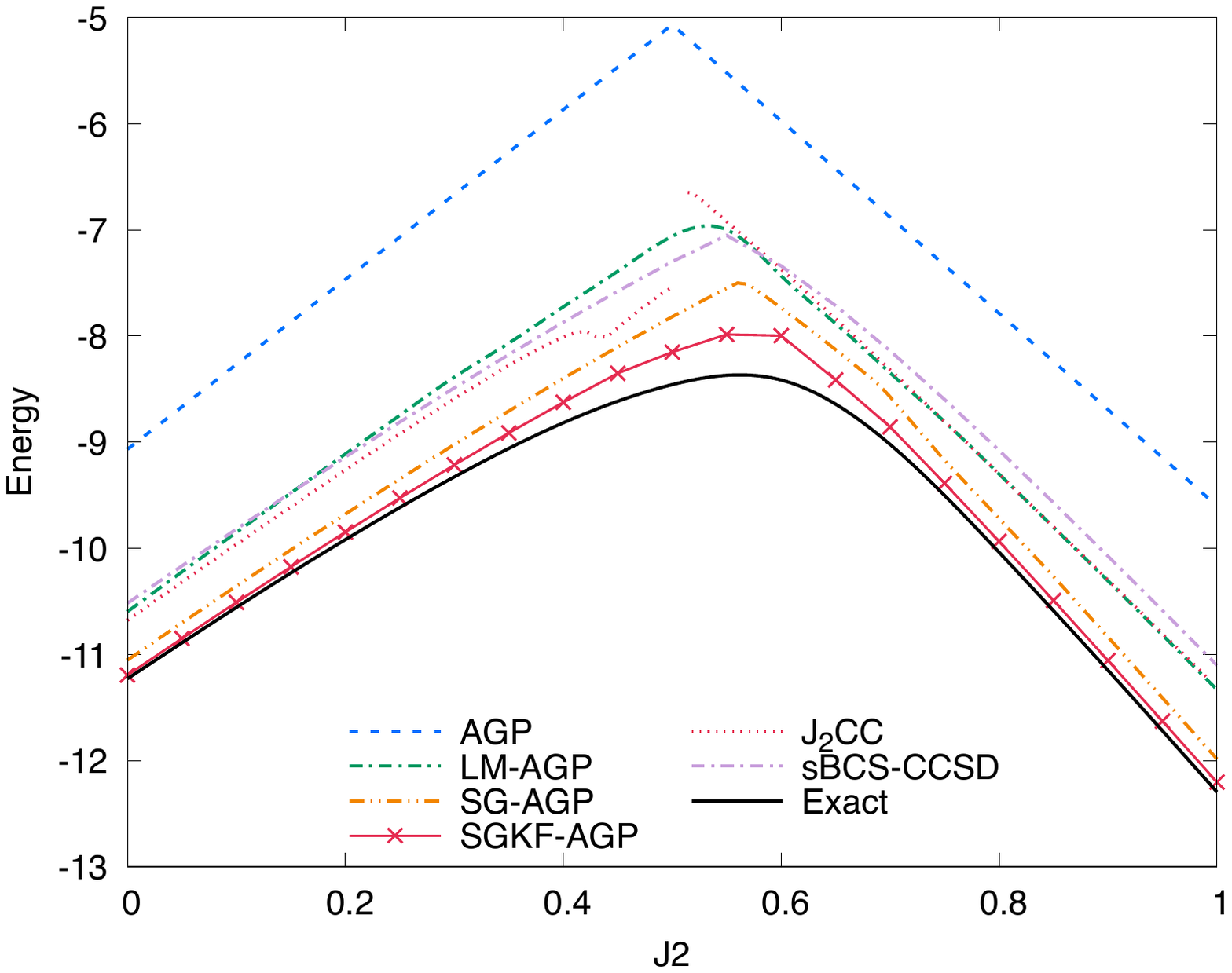}
    \caption{Energy of the $4\times4$ square $\mathrm{J_1-J_2}$ model by space group projected AGP methods compared with mean-field optimized AGP and J$_2$CC and sBCS-CCSD.}
    \label{fig:E_2Dsuqre}
\end{figure}

\begin{figure}[b]
\includegraphics[width=\columnwidth]{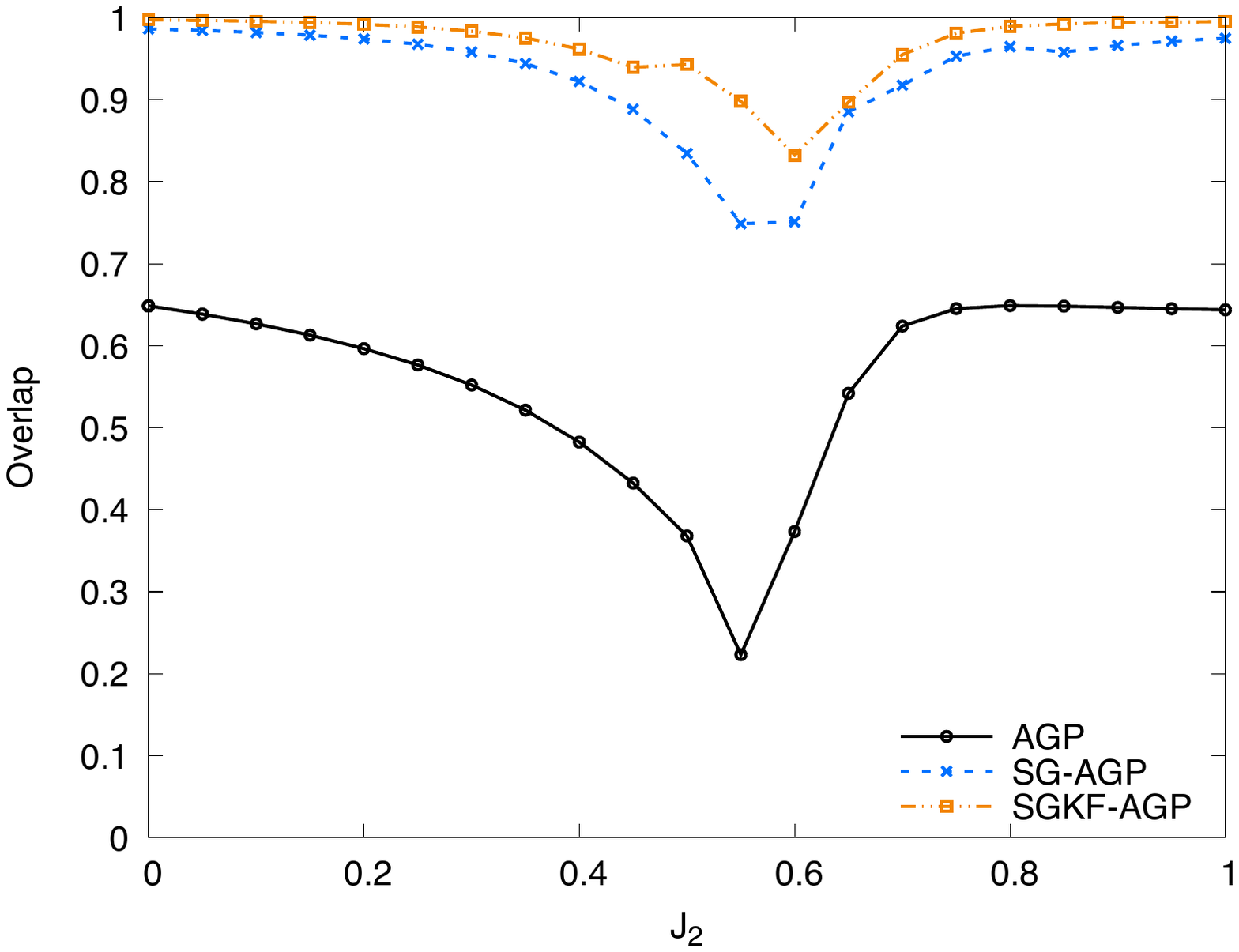}
\caption{AGP, SG-AGP and SGKF-AGP overlap with full CI (FCI).}
\label{fig:2Dsqr_overlap}
\end{figure}

The energy of the $4\times4$ square $\mathrm{J_1-J_2}$ model given by symmetry-projected AGP methods is shown in Fig.~\ref{fig:E_2Dsuqre}. We remind our readers that these energies are optimized with complex $\eta$ via the variation after projection approach. In addition to SG-AGP, we include the results from the combination of space group projection with complex conjugate and spin-flip projection, referred to as SGKF-AGP. Figure~\ref{fig:2Dsqr_overlap} demonstrates their overlap with the exact solution. Both SG-AGP and SGKF-AGP perform well in both the large and small $\mathrm{J_2}$ limits. In the most frustrated region around $\mathrm{J_2}\approx0.5$, they maintain a respectable overlap with the exact solution and succeed in capturing a significant portion of the correlation energy. The energy results are compared with J$_2$CC, which, despite the discontinuity in the highly frustrated region around $\Delta \sim 0.5$, provides some of our best results for correlation atop a single AGP. We also compare the results to spin BCS--based coupled cluster with single and double excitations (sBCS-CCSD), which is the standard approach for treating pair correlations, to offer readers a broader perspective on the accuracy of our methods. We see that with the symmetry projections, SG-AGP and SGKF-AGP can already surpass the performance of correlated methods based on a single AGP or spin BCS.

\begin{figure}
    \centering
    \includegraphics[width=1\linewidth]{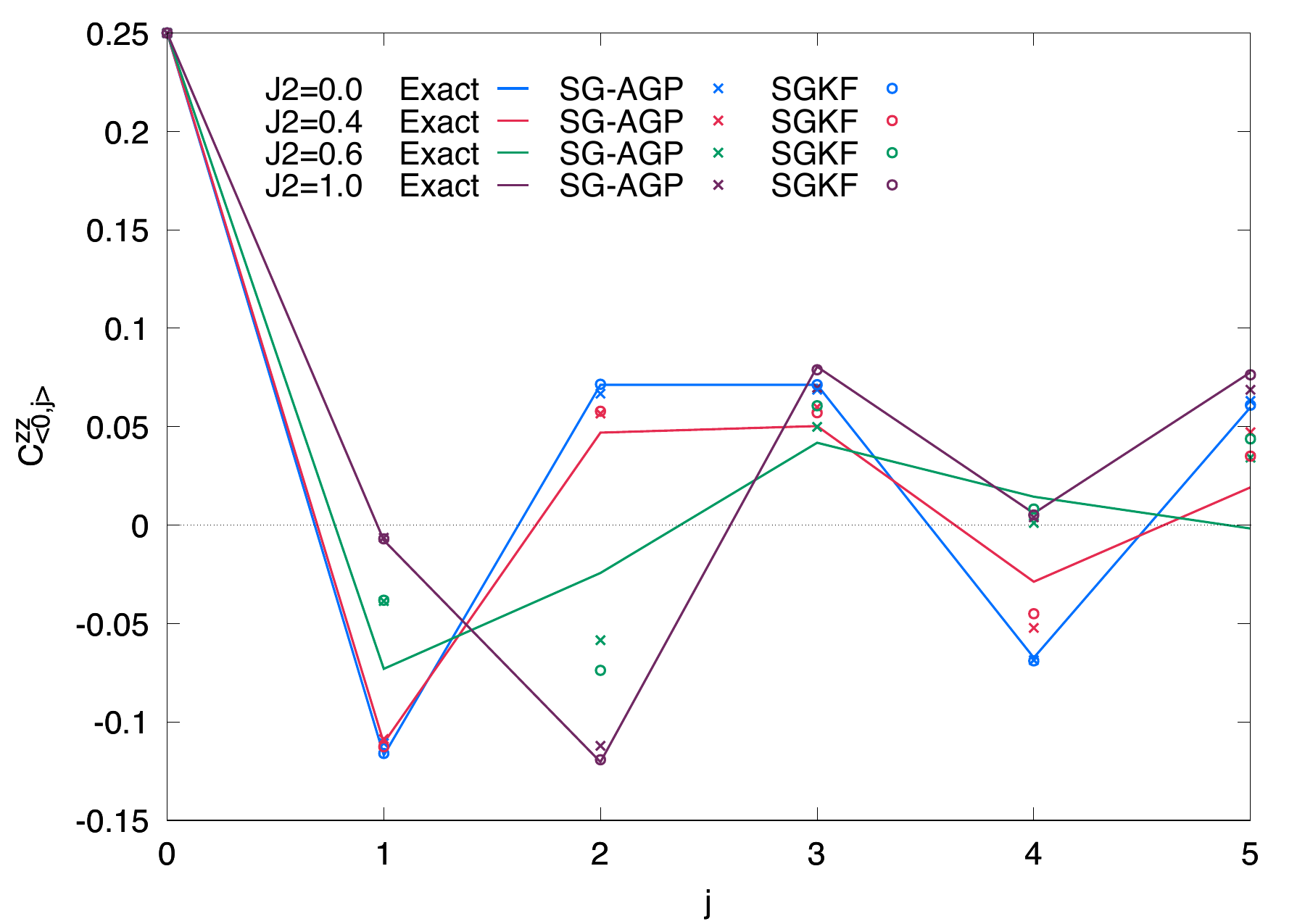}
    
    \vspace{-2.2in}
    
    \hspace{2.2 in}
    \includegraphics[width=0.2\linewidth]{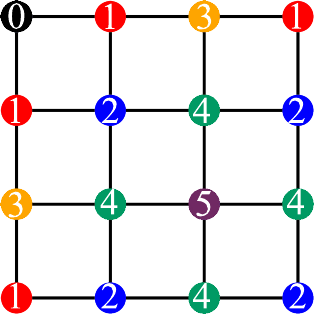}\\
    \vspace{1.4in}
    \centering
    
    \caption{$S^z$-$S^z$ correlation functions from various projected methods in the $4\times4$ square $\mathrm{J_1-J_2}$ model with PBC, compared with exact results. Lines are for the exact result, crosses are for SG-AGP, and circles are for SGKF-AGP. Different colors represent different $\mathrm{J_2}$ values. The horizontal axis $j$ is arranged according to the distance between two spins where $j$ can be regarded as the $j$th-nearest-neighbor of site $0$, as shown in the top right of the figure, where site 0 is colored black and 1---5 are the $j$th-nearest-neighbors, colored differently according to $j$.}
    \label{fig:16squareCorr}
\end{figure}

Figure~\ref{fig:16squareCorr} shows the spin-spin correlation $C^{zz}_{<i,j>} = {\langle S_i^zS_j^z \rangle}$ of the $4\times4$ square $\mathrm{J_1-J_2}$ model. Since the model has translational symmetry and point group symmetry, the correlation $\langle S_i^zS_j^z \rangle$ is only based on the distance between sites $i$ and $j$. In this figure, the indices of the sites are arranged such that site $j$ is the $j$th-nearest-neighbor of site $0$ as shown in the lattice on the top right of the figure.

We observe that when $\mathrm{J_2} = 0$, there is a negative spin-spin correlation for the nearest neighbors, while the second-nearest neighbors exhibit a positive correlation. This is indicative of the checkerboard N\'eel phase. For $\mathrm{J_2} = 1$, the second-nearest-neighbor correlation becomes negative while that of the nearest-neighbor tends toward $0$, corresponding to the striped order. The projected AGP spin-spin correlation functions coincide well with the exact one for large and small $\mathrm{J_2}$. In the spin liquid phase with intermediate $\mathrm{J_2}$ values, the spin-spin correlation is qualitatively correct but shows some deviations for certain site distances.

\subsubsection{Triangular lattice}
\label{Tri_result}
The triangular antiferromagnetic Heisenberg model with nearest-neighbor interaction was the first candidate proposed by P. W. Anderson for a spin liquid phase \cite{anderson1973resonating}, though it was later found to have a 120\textdegree~N\'eel long-range order \cite{PhysRevB.45.12377}. However, when the second-nearest-neighbor coupling ($\mathrm{J_2}$) is included, the triangular $\mathrm{J_1-J_2}$ model indeed exhibits a spin liquid phase around the $\mathrm{J_2/J_1} \approx 0.1$ region \cite{doi:10.7566/JPSJ.83.093707,PhysRevB.91.081104,PhysRevB.92.041105,PhysRevB.92.140403,PhysRevB.93.144411,PhysRevB.94.121111,PhysRevB.96.165141,PhysRevLett.111.157203}. Although the nature of the spin liquid phase is still under debate, studying the spin liquid phase of the $\mathrm{J_1-J_2}$ model may help to understand triangular materials, including $\mathrm{YbMgGaO_4}$ and $\mathrm{Ba_3InIr_2O_9}$.

The spin liquid phase is sandwiched by the 120\textdegree~N\'eel phase for small $\mathrm{J_2/J_1}$ and a stripe N\'eel phase for large $\mathrm{J_2/J_1}$. The actual window of the spin liquid phase is under debate where a coupled-cluster method(CCM) \cite{PhysRevB.91.081104} study predicted it to be at $0.060\lesssim  \mathrm{J_2/J_1}\lesssim 0.165$, while the density matrix renormalization group (DMRG) \cite{PhysRevB.92.041105,PhysRevB.92.140403} suggests $0.08\lesssim  \mathrm{J_2/J_1}\lesssim 0.16$. DMRG found the spin liquid to be a gapped $\mathbb{Z}_2$ spin liquid \cite{PhysRevB.92.041105,PhysRevB.92.140403,PhysRevB.94.121111} and the result is backed by Schwinger-boson mean-field theory \cite{PhysRevB.96.165141} while the CCM \cite{PhysRevB.91.081104} and variational Monte Carlo \cite{doi:10.7566/JPSJ.83.093707} indicate its to be a gapless $U(1)$ Dirac spin liquid.

A typical 2D triangular lattice has a $D_6$ point group symmetry with 12 symmetry operations. For the same point symmetry to be obeyed, a finite-size lattice must be chosen to take the shape of either a rhombus or hexagon, as shown in Fig.~\ref{Fig:twoDLat}. Lattice momentum symmetry involves translation along the 2 lattice vectors. The $D_6$ point group symmetry and the translation symmetry form the $p6m$ space group for the 2D triangular lattice and the SG-AGP for an $M$-site lattice is a linear combination of $12M$ AGP states.

It is worth mentioning that in order to correctly capture the 120\textdegree~phase, the lattice must have a multiple of 3 sites along each translation vector. We will test our methods on the 12-site and 36-site lattices, which satisfy this restriction, and we will also check its behavior on the 16-site lattice, which breaks this restriction as shown in the bottom left of Fig.~\ref{Fig:twoDLat}: if different colors represent spins spaced by 120\textdegree, adjacent sites across the periodic boundary will showcase the same color, thereby breaching the 120\textdegree~phase.

\begin{figure}
    \includegraphics[width=\columnwidth]{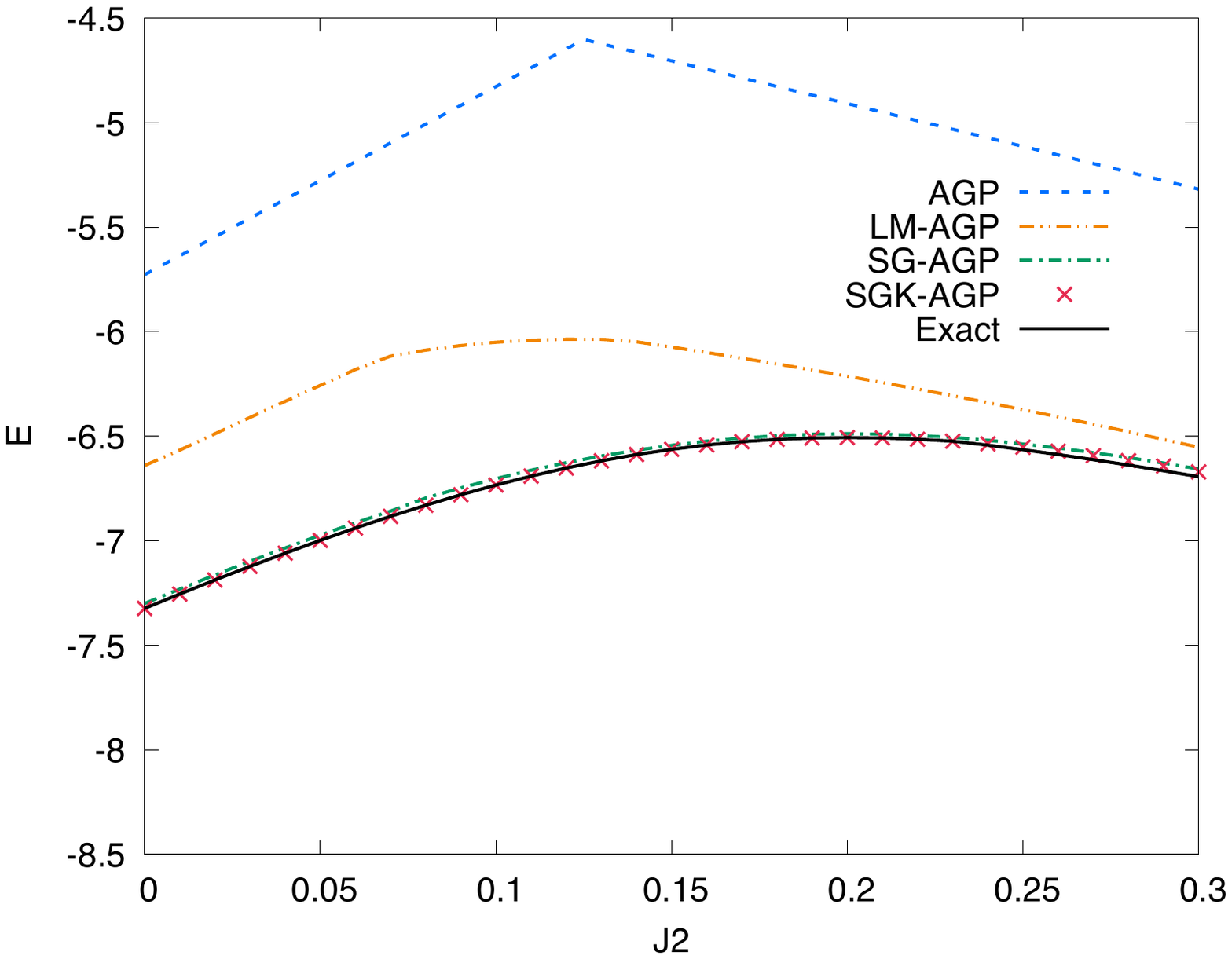}
    \caption{Energy of the 12-site triangular $\mathrm{J_1-J_2}$ model by space group projected AGP methods compared with mean-field optimized AGP. The region of $0.2<\mathrm{J_2}<0.26$ is zoomed in at the bottom right of the figure.}
    \vspace{-2.1in}
    
    \hspace{1.7 in}
    \includegraphics[width=0.38\linewidth]{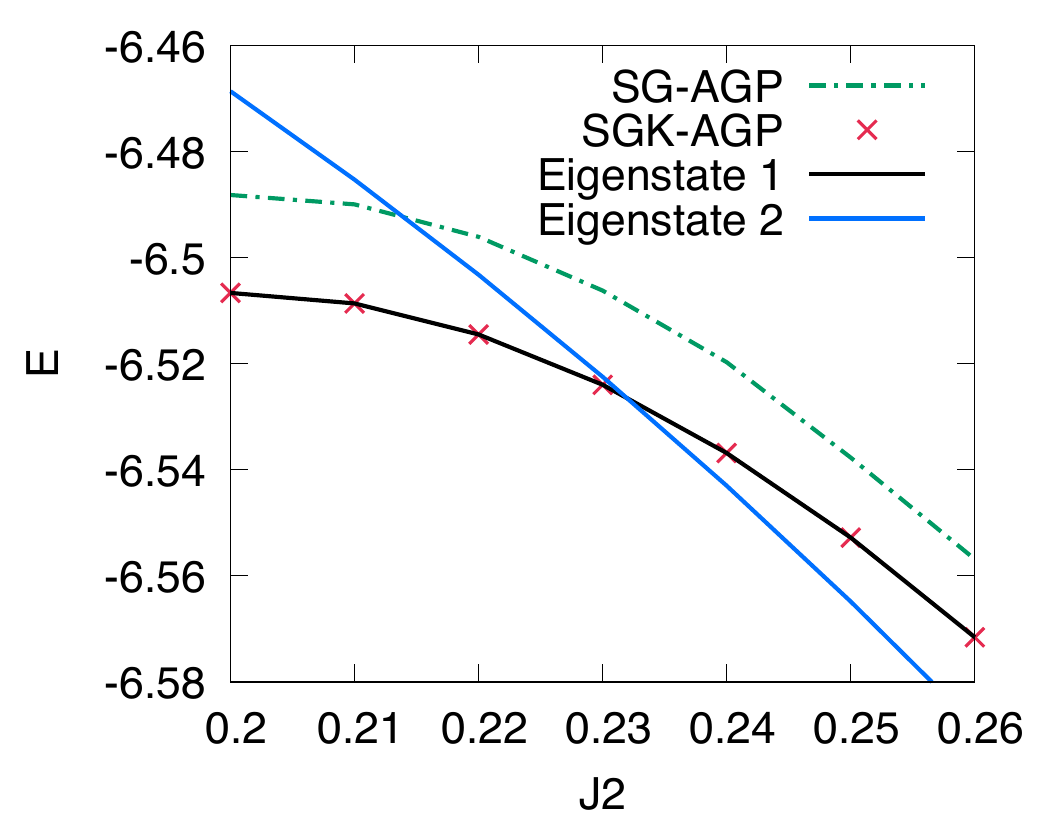}\\
    \vspace{0.8in}
    \centering
    \label{fig:E_2Dtri12}
\end{figure}

\begin{figure}
    \centering
    \includegraphics[width=1\linewidth]{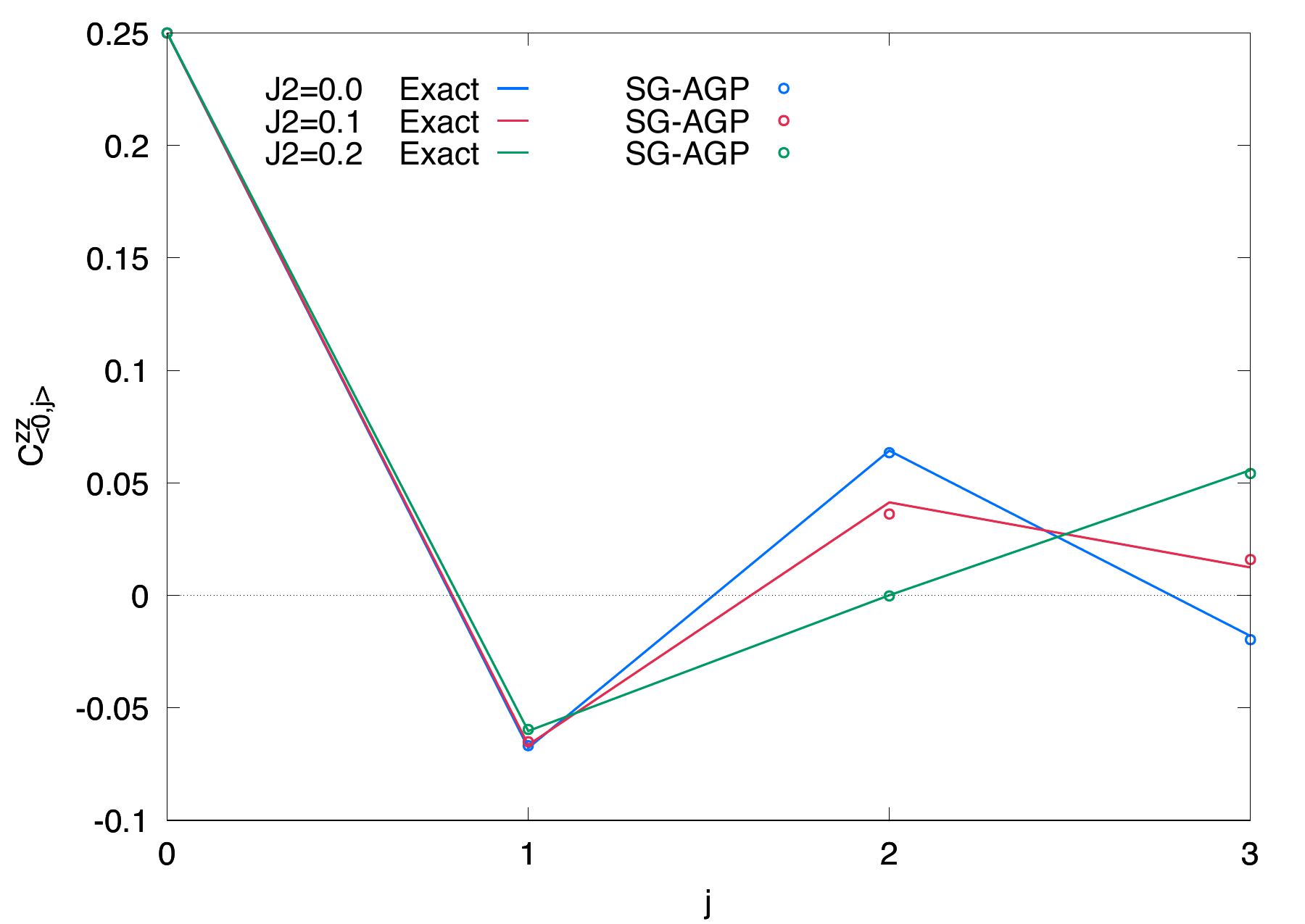}
    
    \vspace{-2.2in}
    
    \hspace{2.0in}
    \includegraphics[width=0.25\linewidth]{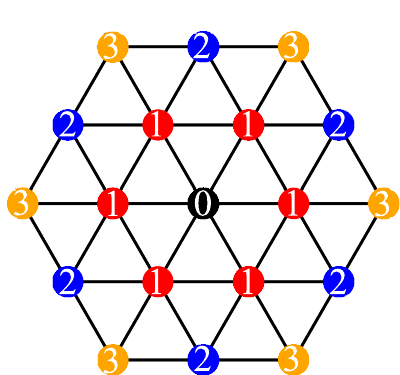}\\
    \vspace{1.4in}
    \centering
    
    \caption{$S^z$-$S^z$ correlation functions of SG-AGP for the 12-site triangular $\mathrm{J_1-J_2}$ model compared with exact results. The lines are for the exact results, and the circles are for SG-AGP. Different colors represent different $\mathrm{J_2}$ values. The horizontal axis $j$ is arranged according to the distance between two spins where $j$ can be regarded as the $j$th-nearest-neighbor of site $0$, as shown in the top right of the figure, where site 0 is colored black and 1---3 are the $j$th-nearest-neighbors, colored differently according to $j$.}
    \label{fig:12triCorr}
\end{figure}

The 12-site lattice has a hexagonal shape, as shown in the top right of Fig.~\ref{Fig:twoDLat}. The energy of space group projected AGP methods is shown in Fig.~\ref{fig:E_2Dtri12}. Although a single mean-field optimized AGP provides only a qualitative insight, including additional symmetry projections brings the energy closer to the exact value, with the SG-AGP exhibiting an error of $10^{-2}$. Moreover,  one of the exact eigenstates is obtained when the complex conjugate projection is included, which might be attributed to the limited size of the system. As shown in the bottom right of the figure, the SGK-AGP and the eigenstate feature the ground state for $\mathrm{J_2}<0.23$, after which a different eigenstate with a different spatial symmetry has a lower energy and SGK-AGP is no longer the ground state but still an eigenstate of the system. The spin-spin correlation of SG-AGP for this lattice is illustrated in Fig.~\ref{fig:12triCorr}, demonstrating that SG-AGP accurately captures the correlation.

\begin{figure}
    \includegraphics[width=\columnwidth]{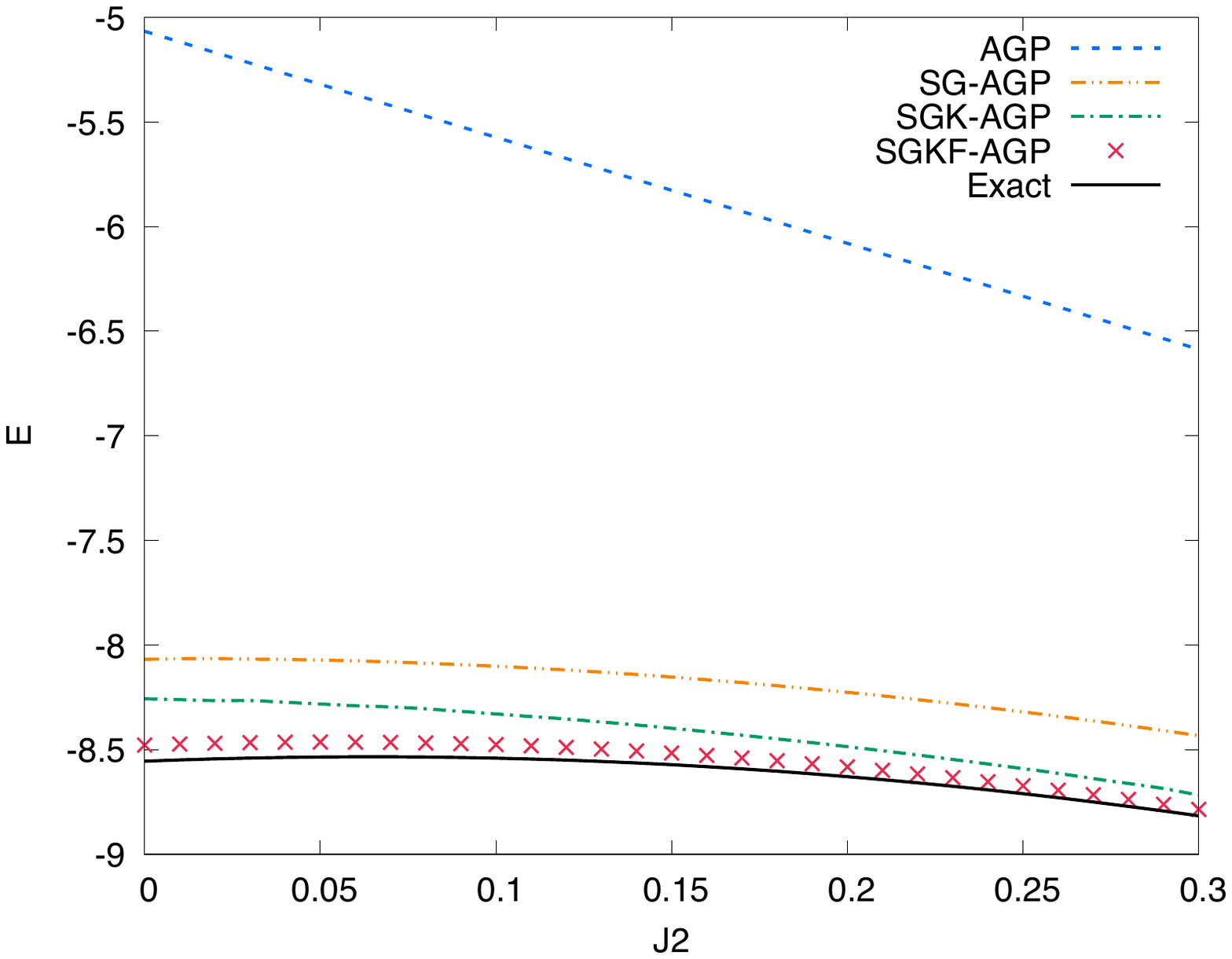}
    \caption{Energy of the 16-site triangular $\mathrm{J_1-J_2}$ model by space group projected AGP methods compared with mean-field optimized AGP. Due to the finite-size effect, the energy curves differ from those of the 12-site and 36-site lattices.}
    \label{fig:E_2Dtri16}
\end{figure}

\begin{figure}
   \includegraphics[width=\columnwidth]{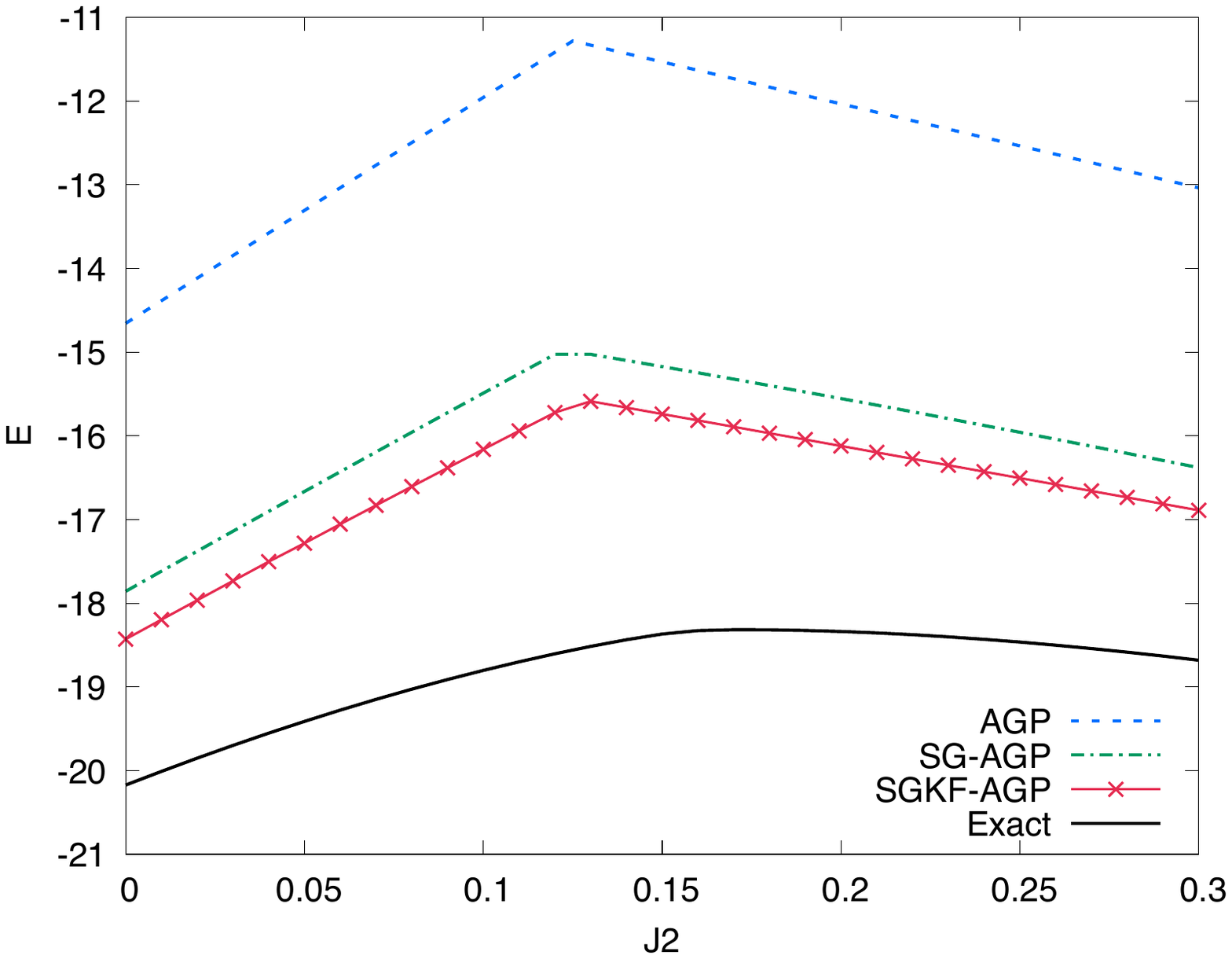}
    \caption{Energy of the 36-site triangular $\mathrm{J_1-J_2}$ model by space group projected AGP methods compared with mean-field optimized AGP.}
    \label{fig:E_2Dtri36}
\end{figure}

The space group projected AGP methods are also tested on the 16-site ($4\times4$ rhombus) and 36-site ($6\times6$ rhombus) triangular lattices shown in Fig.~\ref{fig:E_2Dtri16} and Fig.~\ref{fig:E_2Dtri36} respectively. In the 16-site lattice, where the lattice size is not a multiple of 3, due to the finite-size effect mentioned earlier, the energy curves of both the symmetry-projected AGP methods and the exact solution deviate from those observed in the 12-site and 36-site lattice models. Despite the absence of the 120\textdegree~phase, the exact ground state of this lattice continues to exhibit the same space group symmetry as other system sizes. The space group projected AGP methods can accurately capture the energy, particularly SGKF-AGP, which has an error of approximately $10^{-1}$.

\begin{figure}[b]
    \centering
    \includegraphics[width=1\linewidth]{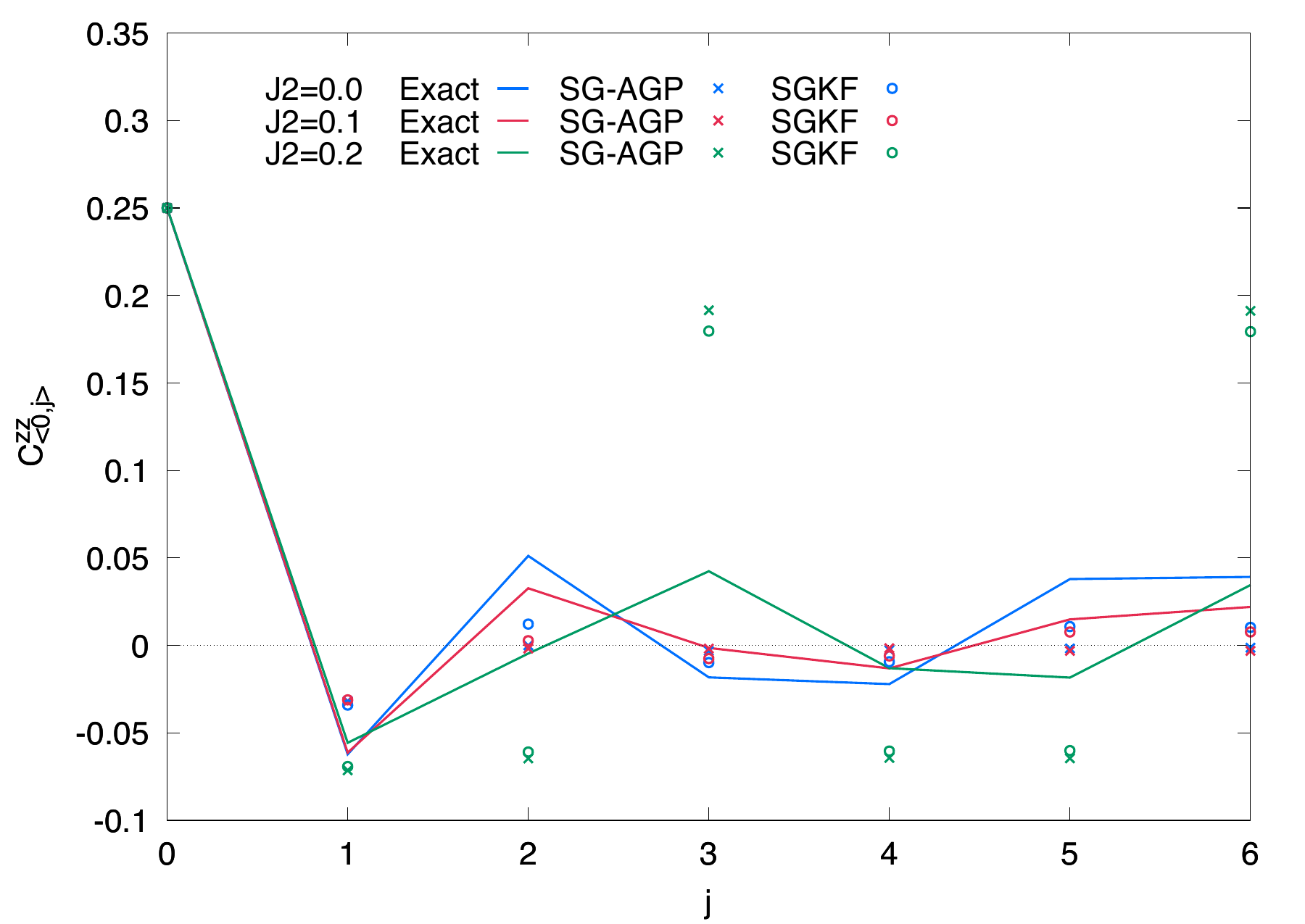}

    \caption{$S^z$-$S^z$ correlation functions from various projected methods in the 36-site triangular $\mathrm{J_1-J_2}$ model with PBC, compared with exact results. Lines are for the exact result, crosses are for SG-AGP, and circles are for SGKF-AGP. Different colors represent different $\mathrm{J_2}$ values. The horizontal axis $j$ is arranged according to the distance between two spins where $j$ can be regarded as the $j$th-nearest-neighbor of site $0$.}
    \label{fig:36triCorr}
\end{figure}

However, an important limitation of projected AGP methods (and of AGP itself) is that they are not size extensive.  In the limit of large lattices, the energy per site reduces to that of the spin-BCS.  Figure~\ref{fig:E_2Dtri36} and Figure~\ref{fig:36triCorr} shows energy and spin-spin correlation results for the 36-site rhombus lattice. We can see that the SG-AGP energy is significantly less accurate here than it is for the 12-site hexagon. The spin-spin correlation for the SGKF-AGP retains the correct sign, but it is not quantitatively accurate. The shortcomings for SG-AGP must at least in part be due to the lack of extensivity, though it may also be related to the differently-shaped lattice.  Unfortunately, the next larger hexagonal lattice contains 48 sites and is too large for us to generate exact results against which we can compare, and there is no smaller rhombus that is both compatible with the 120\textdegree~phase and possesses an $S^z=0$ ground state. As a result, we cannot generate exact data to discriminate between effects due to lattice \textit{size} and those due to lattice \textit{shape}.

Regardless of whether the lower accuracy seen in Fig.~\ref{fig:E_2Dtri36} is simply due to lack of extensivity or not, it is clear that projected AGP methods are inadequate in the thermodynamic limit. To address larger systems accurately, it is imperative to integrate correlated approaches, including J$_2$CC, with the symmetry projection methods.  This requires further research. Alternatively, due to the superiority of symmetry-projected methods on small and medium-sized systems, it is also feasible to incorporate space group projection techniques with cluster mean field \cite{doi:10.1021/acs.jctc.2c00338,10.1063/5.0155765} or density matrix embedding theory \cite{PhysRevB.89.035140}, where large systems are composed of fragments and SG-AGP may provide a good description of the fragments.

\section{Conclusions}

In this article, we have studied symmetry-projected AGP methods for the 1D XXZ model and the 2D $\mathrm{J_1-J_2}$ model with square and triangular lattices. The projection of space group symmetry is achieved with a linear combination of permuted AGPs, resulting in a computational cost of $\mathcal{O}(M^4)$, which is one factor of $M$ more expensive than a single mean field optimized AGP. The symmetry-projected AGP is optimized with the variation after projection approach. It can be seen that the space group projected AGP not only restores the correct symmetry of the system but also adds important strong correlations that a single optimized AGP misses.

Space-group projected AGP methods work better when we use a wavefunction that breaks complex conjugation symmetry. This symmetry can be restored with a further complex-conjugation symmetry projection upon the space group projected AGP. In addition, spin-flip symmetry and time-reversal symmetry are also discussed in this paper. The three symmetries form a closed set, and they serve as a significant complement to the space group projection for the symmetry projection. Combining these projections together, the symmetry-projected AGP methods behave well both for the energy and spin-spin correlation for the 1D XXZ lattice and 2D square $\mathrm{J_1-J_2}$ model and can outperform single-AGP referenced correlated methods.

As with all symmetry-projected methods, the quality of symmetry-projected AGP deteriorates as the system size increases.  We note, however, that symmetry-projected AGP is essentially a mean-field method, and significantly improved results may be achieved by combining symmetry-projected AGP with methods intended to incorporate further correlations \cite{Duguet_2015,Duguet_2017,doi:10.1021/acs.jctc.7b00073,10.1063/1.5036542,PhysRevC.99.044301,10.1063/1.4991020,10.1063/1.5053605,10.1063/5.0080165}.

\begin{acknowledgments}

This work was supported by the U.S. Department of Energy, Office of Basic Energy Sciences, Computational and Theoretical Chemistry Program under Award DE-SC0001474. G.E.S. acknowledges support as a Welch Foundation Chair (Grant No. C-0036). We thank Dr. Ruiheng Song for generously providing the sBCS-CCSD results shown in Fig.~\ref{fig:E_2Dsuqre}.

\end{acknowledgments}

\bibliographystyle{apsrev4-2}
\bibliography{SGAGP}

\end{document}